\documentstyle[12pt]{article}
\begin{document}
\everymath={\displaystyle}
\thispagestyle{empty}
\noindent
\hfill TTP93--14\\
\mbox{}
\hfill  May  1993   \\   %                 \hfill \today \\
\vspace{0.5cm}
\begin{center}
  \begin{Large}
  \begin{bf}
 TAU KINEMATICS FROM IMPACT PARAMETERS
 \footnote{\normalsize Supported by BMFT Contract 056KA93P}
   \\
  \end{bf}
  \end{Large}
  \vspace{0.8cm}
  \begin{large}
   J.H. K\"uhn \\[5mm]
    Institut f\"ur Theoretische Teilchenphysik\\
    Universit\"at Karlsruhe\\
    Kaiserstr. 12,    Postfach 6980\\[2mm]
    7500 Karlsruhe 1, Germany\\
  \end{large}
  \vspace{4.5cm}
  {\bf Abstract}
\end{center}
\begin{quotation}
\noindent
The momenta of $\tau$ decay products in the reaction
$e^+e^-\to\tau^+\tau^-$ do
not constrain the $\tau$ direction unambiguously.  It is shown how
the measurement of tracks of hadrons from semileptonic $\tau$
decays, in particular their relative impact parameters, allows to
resolve this ambiguity.
\smallskip
\end{quotation}
%++++++++++++++++++++++macros++++++++++++++++++++++++++++++++++++++++
\newcommand\beq{\begin{equation}}
\newcommand\eeq{\end{equation}}
\newcommand\tlp{\theta^L_+}
\newcommand\tlm{\theta^L_-}
\newcommand\cp{\cos\tlp}
\newcommand\cm{\cos\tlm}
\newcommand\spp{\sin\tlp}
\newcommand\sm{\sin\tlm}
\newcommand\np{\vec n_+}
\newcommand\nm{\vec n_-}
\newcommand\Pp{\vec p_+}
\newcommand\Pm{\vec p_-}
\newcommand\cph{\cos\varphi}
\newcommand\sph{\sin\varphi}
\newcommand\vd{\vec d}
\newcommand\dmin{\vd_{min}}
%+++++++++++++++++++++end macros+++++++++++++++++++++++++++++++++++++
%-----------------------------------------------------------------------
\newpage
\setcounter{page}{1}

The analysis of $\tau$ pair production in electron-positron
annihilation has advanced considerably during the past years as a
consequence of increasing statistics combined with improved
detector performance \cite{Drell}.  $\tau$ decays into hadrons were
particularly useful.  On the one hand they allow for
determination of the $\tau$ coupling to the (virtual) $W$ and the $Z$,
on the other hand they lead to a measurement of the structure
functions \cite{KM1,KM2} and subsequently of the form factors.
However, the kinematic configuration of these events cannot be fully
reconstructed since the momenta of both $\nu_\tau$ and $\bar\nu_\tau$
are unknown.
The resulting loss of information can be recovered only partially by
measuring appropriately chosen energy, momentum, and angular
distributions \cite{KW}--
%,Hag,Roug,KM1,KM2,Dav,
\cite{Priv}.

In events where both leptons decay semileptonically and where all
hadron momenta are determined, the original $\tau$ direction can be
reconstructed up to a twofold ambiguity \cite{Tsai,KW,Dav}.
In this paper we show
that this can be resolved with the help of vertex detectors
employed in present experiments.  Several possibilities may
arise:

{\em i.)}  If the beam spot is large compared to the typical impact
parameter, the production vertex is unknown.  Let us assume that
both $\tau$ decay into one charged hadron each and that both
charged tracks can be measured with high precision.  The
direction $\vec d_{min}$
of the minimal distance between the two nonintersecting
charged tracks (Fig.1) resolves the ambiguity and introduces two
additional constraints that can be used to reduce the measurement
errors.  The $\tau^+$ and $\tau^-$ decay points and their original
direction of flight are then uniquely determined.

{\em ii.)}  Precise knowledge of the beam axis (corresponding to a
beam spot of negligible size) leads to a further constraint
resulting from the requirement that the reconstructed $\tau$ axis
and the beam axis intersect.  If the production point would be
known in addition then the momenta plus one charged track from the decay
of only one $\tau$ would allow to reconstruct the event
and double semileptonic decays
would lead to a large number of additional constraints.

{\em iii.)}  If one (or both) $\tau$ decay into several hadrons and
if all momenta and two tracks (one from each side) are measured,
the same reconstruction can be performed.

Let us first consider case {\em i)} where all relevant aspects can be
explained most clearly.  The angles $\theta^L_{\pm}$ between the
$\tau^\pm$ and the  hadron $h^\pm$
directions respectively as defined in the lab frame are
given by the energies of $h^+$ and $h^-$ \cite{KW}:
\beq
\cm=\frac{\gamma x_- - (1+r^2_-)/2\gamma}
                      {\beta\sqrt{\gamma^2x_-^2-r_-^2}}
\eeq
\beq
\sm=
    \sqrt{\frac{(1-r_-^2)^2/4 - (x_--(1+r^2_-)/2)^2/\beta^2}
                      {\gamma^2x_-^2-r_-^2}}
\eeq
\beq
x_-=E_{h^-}/E_\tau \hspace{2cm}   r_-=m_{h^-}/m_\tau
\eeq
and similarly for $\cp$ and $sp$. The velocity
$\beta$, and the boost factor $\gamma$ refer to the $\tau$ in the lab
frame.

The original $\tau^-$ direction must therefore lie on the cone of
opening angle $\tlm$ around the direction of $h^-$ and on the cone of
opening angle $\tlp$ around the reflected direction of $h^+$.  The
extremal situation where $\tlp$ or $\tlm$ assume the values 0 or $\pi$,
or where the
two cones touch in one line, lead to a unique solution for the $\tau$
direction.  In general a twofold ambiguity arises as
is obvious from this geometric argument. The
cosine of the relative azimuthal angle $\varphi$ between the directions
of $h^+$ and $h^-$ denoted by $\np$
and $\nm$ can be calculated from the momenta
and energies of $h^+$ and $h^-$ as follows:  In the coordinate frame
(see Fig.1) with the $z$ axis pointing along the direction of $\tau^-$
and with $\nm$ in the $xz$ plane and positive $x$ component
\beq
\frac{\Pm}{|\Pm|}\equiv\nm=
\left(\begin{array}{c} \sm    \\0       \\\cm     \end{array}\right)
\hspace{1cm}
\frac{\Pp}{|\Pp|}\equiv\np=
\left(\begin{array}{c}\spp\cph \\ \spp\sph\\-\cp    \end{array}\right)
\eeq
and $\cph$ can be determined from
\beq
\nm\np=-\cm \cp + \sm \spp\cph
\eeq
The well-known twofold ambiguity in $\varphi$ is evident from this formula.

Additional information can be drawn from the precise
determination of tracks close to the production point.
Three-prong decays allow to reconstruct the decay vertex and the
ambiguity can be trivially resolved.

However, also single-prong events may serve this purpose.  Let us
first consider decays into one charged hadron on each side.
Their tracks and in particular the vector $\dmin$ of closest
approach
(Fig.1) can be measured with the help of microvertex detectors.
The vector pointing
from the $\tau_-$  to the $\tau_+$ decay vertex
\beq
\vec d\equiv \vec \tau_+ -\vec\tau_-= -l
\left(\begin{array}{c}0\\0\\1\end{array}\right)
\eeq
is oriented  by definition into the negative $z$ direction
($l>0)$.
The vector $\dmin$ can on the one hand be measured, on the other hand
calculated from $\vec d$, $\np$ and $\nm$:
\beq\label{eqdmin}
\dmin=\vd\, + \,
[(\vd\np\,\np\nm - \vd\nm)\nm + (\vd\nm\,\np\nm - \vd\np)\np]
/(1-(\nm\np)^2)
\eeq
The sign of the projection of $\dmin$ on $\np\times\nm$ then
determines the sign of $\varphi$ and hence resolves the
ambiguity.
\beq
\dmin(\np\times\nm) = l \spp\sm\sph
\eeq
The length of the projection determines $l$ and hence
provides a measurement of the lifetimes
of
$\tau_+$ plus $\tau_-$.
Exploiting the fact that $\vd\nm=-l\cm$ and $\vd\np=l\cp$ the direction
of $\vd$ can be geometrically constructed by inverting (\ref{eqdmin}):
\beq
\vd/l=\dmin/l \, - \,
[(\cp \, \np\nm + \cm)\nm + (-\cm \, \np\nm - \cp)\np] / (1-(\nm\np)^2)
\eeq
In addition two constraint equations may be
derived by comparing $\dmin/l$
as calculated from the $h_+$ and $h_-$ tracks with
the direction calculated from (\ref{eqdmin})
with the help of $\tlp$, $\tlm$ and $l$.
These might be used to constrain the events even in cases where initial
state radiation distorts the simple kinematics described above.

As stated before, the locations of both $\tau_+$ and $\tau_-$
decay vertices
in space are then fixed.  If the beam axis is known with high
precision (high compared to the decay length $l$) the lines between
the two decay vertices and the beam axis intersect, providing one
additional constraint.

The generalization of this method to decays into multihadron
states with one or several neutrals is straightforward: $\tlp$,
$\tlm$ and $\cos\varphi$ are
fixed by the hadron momenta as stated above.  Only one of the two
solutions for $\varphi$ is then compatible with $\dmin$
measured directly with the help of vertex detectors.

In summary:  the measurement of hadron momenta in events of the
type \\ $e^+e^-\to\tau^-(\to h^-\nu)\tau^+(\to h^+\bar\nu)$
combined with an accurate determination of even one charged
track on either side allows for a full reconstruction of the $\tau$
direction.  A full analysis of angular distributions in
semileptonic decays involving $\tau$ and hadron momenta
as discussed in the appendix of \cite{KM2}
therefore appears feasible.
%-----------------------------------------------------------------------
\sloppy
\raggedright

%==============================================================
%\newpage
\begin{figure}[htb]
\vspace{2.cm}
\caption{}
Kinematic configuration indicating the relative orientation of the
hadronic tracks, the $\tau$ directions and the vector $\dmin$.
\label{Kin}
 \end{figure}
\end{document}